hep-lat/9311064  1 Dec 1993

# Euclidean gravity attracts[*]

Bas de Bakker and Jan Smit

Institute for Theoretical Physics, University of Amsterdam,
Valckenierstraat 65, 1018 XE Amsterdam, the Netherlands

We look at gravitational attraction in simplicial gravity using the dynamical triangulation method. On the dynamical triangulation configurations we measure quenched propagators of a free massive scalar field. The masses measured from these propagators show that gravitational attraction is present.

## 1. Introduction

One of the approaches to quantum gravity that has recently raised increasing interest is that of dynamical triangulation. More details on this method by itself can be found in e.g. [1,2].

The one thing that everybody knows about gravity is that it is an attractive force. This led us to the wish of actually seeing some of this attraction in dynamical triangulation.

## 2. Description of the model

We look at the behaviour of a free scalar field $\phi$ with bare mass $m_0$ in a quantum gravity background. The euclidean action of this system in continuum language would be

$$S = S[g] + S[g, \phi], \tag{1}$$

$$S[g] = \frac{1}{16\pi G_0} \int d^D x \sqrt{g} \, (2\Lambda_0 - R), \tag{2}$$

$$S[g, \phi] = \int d^D x \sqrt{g} \times \left( \frac{1}{2} g^{\mu\nu} \partial_\mu \phi \partial_\nu \phi + \frac{1}{2} m_0^2 \phi^2 \right), \tag{3}$$

where $\Lambda_0$ is the bare cosmological constant, $R$ is the scalar curvature and $G_0$ is the bare Newton constant.

We take $\phi$ as a test particle here, i.e. the back reaction of the field $\phi$ on the metric is not taken into account. It is seen in other simulations [3] that such matter has little influence on the gravity sector of the theory.

We will use the following notation for expectation values of an observable $A$.

$$\langle A \rangle_\phi = \frac{\int \mathcal{D}\phi \, A \exp(-S[g, \phi])}{\int \mathcal{D}\phi \, \exp(-S[g, \phi])}, \tag{4}$$

$$\langle A \rangle_g = \frac{\int \mathcal{D}g \, A \exp(-S[g])}{\int \mathcal{D}g \, \exp(-S[g])}, \tag{5}$$

The quenched expectation value is then

$$\langle A \rangle = \langle \langle A \rangle_\phi \rangle_g. \tag{6}$$

We can now look at propagators in a fixed geometry. The one particle propagator, denoted by $G(x)$, is defined as

$$G(x) = \langle \phi_x \phi_0 \rangle_\phi, \tag{7}$$

where 0 is an arbitrary point. The connected two particle propagator will then be the square of the one particle propagator

$$\langle \phi_x \phi_x \phi_0 \phi_0 \rangle_{\phi, \text{conn}} = G(x)^2. \tag{8}$$

Letting the metric fluctuate, we take the average of the propagators over the different metrics. Because of reparametrization invariance, the average $\langle G(x) \rangle_g$ will not depend on the place $x$. Therefore, we look at averages at fixed geodesic distance $d$

$$\langle G(d) \rangle_g = \frac{1}{Z} \int \mathcal{D}g \, \exp(-S[g]) \times \int d^D x \, G(x) \, \delta(d(x, 0) - d), \tag{9}$$

where $d(x, y)$ is the minimal geodesic distance between $x$ and $y$. For a massive particle, we expect

---

[*]Presented by Bas de Bakker

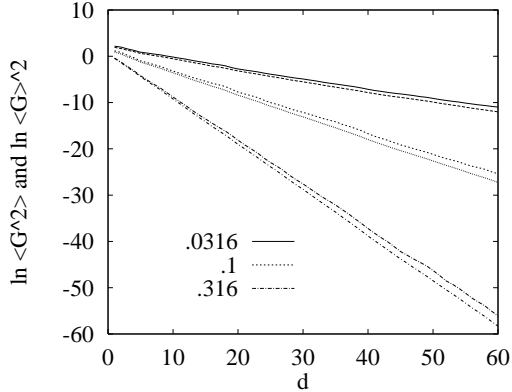

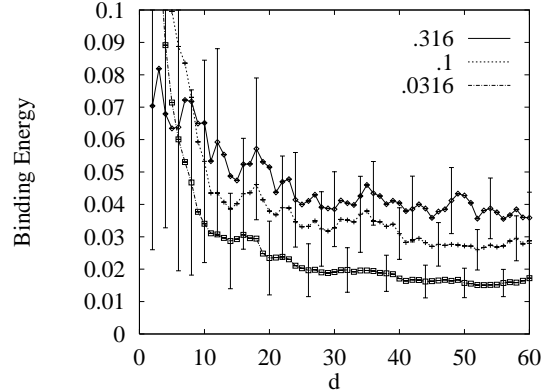

Figure 1. The two particle and the square of the one particle propagator versus the geodesic distance for three different bare masses in two dimensions. The vertical scale is logarithmic.

Figure 2. The effective binding energy $E_b$ as a function of the geodesic distance for three different bare masses in two dimensions.

this propagator to fall off exponentially as

$$\langle G(d) \rangle_g \quad \propto \quad d^\alpha \exp(-md), \qquad (10)$$

$$\langle G(d) \rangle_g^2 \quad \propto \quad d^{2\alpha} \exp(-2md), \qquad (11)$$

with some power $\alpha$ and the measured mass $m$. The two particle propagator will behave similarly as

$$\langle G(d)^2 \rangle_g \quad \propto \quad d^\beta \exp(-E_c d), \qquad (12)$$

where $E_c$ is the energy of the two particle compound. It this energy turns out to be less than two times the mass of a single particle, the difference can be interpreted as a binding energy between the particles. This would show gravitational attraction between them.

## 3. Implementation

We have run numerical simulations with both two and four dimensional dynamical triangulations.

In two dimensions the volume can be kept constant, and for fixed topology no parameters will be left. We used systems of 32k and 64k triangles with the topology of the two-sphere.

In four dimensions the analogue of the continuum gravitational action is

$$S[g] \quad = \quad \frac{1}{16\pi G_0} \int d^4x \sqrt{g}\, (2\Lambda_0 - R)\,, \qquad (13)$$

$$\longrightarrow \quad \kappa_4 N_4 - \kappa_2 N_2, \qquad (14)$$

where $N_2$ and $N_4$ are the number of triangles and four-simplices repsectively. We used systems of about 16k simplices and the topology of the four-sphere. To keep the number of simplices around the desired value, we varied $\kappa_4$, increasing it when the volume became too large and vice versa.

The parameter $\kappa_2$ is inversely proportional to the bare gravitational constant $G_0$. As in other work [1–4], we see indications of a second order phase transition as $\kappa_2$ varies.

On each dynamical triangulation configuration we then calculated the propagator

$$G(x) = (\Box^2 + m_0^2)_{0x}^{-1} \qquad (15)$$

of the scalar field, using the algebraic multigrid routine AMG1R5.

Eventually we hope to be able to extract the renormalized Newton's constant $G_R$ at the critical value of $\kappa_2$, e.g. according to the nonrelativistic formula

$$E_b = \frac{1}{4} G_R^2 m^5\,. \qquad (16)$$

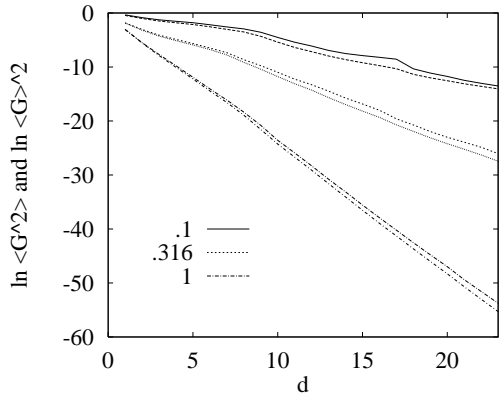

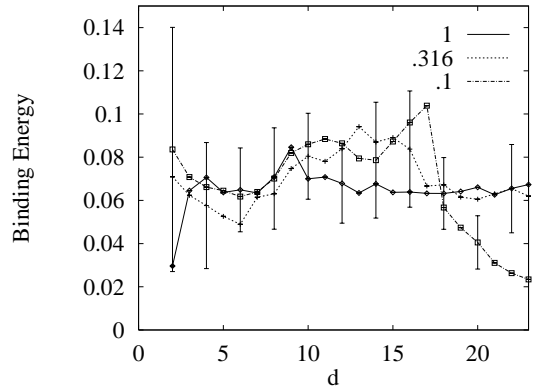

Figure 3. As in figure 1, but in four dimensions. $\kappa_2 = 1.2$, which corresponds to the weak coupling (elongated) phase.

Figure 4. As in figure 2, but in four dimensions. Again, $\kappa_2 = 1.2$.

## 4. Results

In figure 1 we see the results in two dimensions for three different bare masses. Each pair of lines corresponds to one bare mass. In each pair the upper line is $\ln\langle G(d)^2\rangle$ (the two particle propagator) and the lower line is $\ln\langle G(d)\rangle^2$ (the product of two single particle propagators).

Their is clearly a difference in slope between the lines in each pair. This shows that the energy of the two particle compound is less than two times the mass of a single particle and consequently that there is a positive binding energy between the particles.

Figure 3 shows similar data in four dimensions. The propagators at the lowest mass show large fluctuations, but at the higher masses there is again a clear difference in slope.

In the two dimensional case, one might expect not to see any attraction, because of the absence of dynamics in the classical system. In the quantum case, however, a non-trivial theory can be seen due to the conformal anomaly [5].

From these data, we can now calculate the binding energy of the particles. From (11) and (12) we have

$$E_b \equiv 2m - E_c \tag{17}$$

$$= d^{-1}\ln\frac{\langle G(d)^2\rangle_g}{\langle G(d)\rangle_g^2}, \qquad d \to \infty. \tag{18}$$

Figure 2 shows this quantity as a function of the geodesic distance. The three curves again correspond to the three different bare masses. Although the result is not yet very accurate, it is clear (and fits to these curves show) that the binding energy goes to a non-zero value.

Figure 4 shows the same data in the four dimensional simulations. Obviously, these data are very preliminary and we need more statistics. Nevertheless, this figure also indicates a non-zero binding energy.